\documentstyle[prl,aps,epsfig,multicol]{revtex}  

\begin{document}
 
\title{Comment on ``Stability of Small Carbon-Nitride Heterofullerenes"} 
 
\author{Guangyu Sun$^{1}$, Rui-Hua Xie$^{2}$, and Marc C. Nicklaus$^{1}$} 
\address{$^{1}$Laboratory of Medicinal Chemistry, NCI-Frederick, Frederick, MD 21702\\
$^{2}$National Institute of Standards and Technology, Gaithersburg, MD 20899-8423} 
 
\date{\today}

\maketitle

\begin{multicols}{2} 

In a recent Letter \cite{ref1}, Schultz et al. proposed the 
existence of heterofullerenes C$_{n-x}$N$_{x}$ ($40\le  n \le  50$) 
based on the pattern of the peaks in the mass spectrum of 
carbon-nitride nanostructures produced by the arc-discharge technique. 
Using PM3 semiempirical calculations, they predicted the most stable structures of 
the N-substituted fullerenes, C$_{33}$N$_{9}$ and C$_{44}$N$_{6}$, to have 
$C_{3}$ symmetry. While some level of symmetry seems to 
maximize stability, the highest symmetry often sacrifices 
it, as in the case of fullerene C$_{84}$\cite{ref2}. Using PM3 and 
density functional theory (DFT), we argue that a highly systematic investigation 
is needed  to determine the most stable structures of both pristine and N-substituted 
fullerenes. We demonstrate this explicitly for two particular cases, 
C$_{40}$ and C$_{44}$N$_{6}$, for which the authors of 
 Ref.\cite{ref1} only considered four and two isomers, respectively.

Mathematically, there are 40 possible isomers for C$_{40}$\cite{ref3}. 
We optimized the geometries of the 37 isomers with non-vanishing HOMO-LUMO 
gaps using the PM3 Hamiltonian and the B3LYP functional with STO-3G, 3-
21G and 6-31G* basis sets as implemented in Gaussian 98 \cite{ref4}. The 
closed-shell formalism was used and the topological symmetry was maintained throughout. 
Isomers {\bf 1}, {\bf 30} and {\bf 40} (note: the isomer index follows Ref.\cite{ref3})
 have zero HOMO-LUMO gaps at the H\"{u}ckel level of theory. Geometric convergence could 
not be achieved for these cases at the PM3 level and no 
further calculation was performed. 

The most stable isomer of C$_{40}$ is predicted to be isomer 
{\bf 38} ($D_{2}$) by all levels of theory. Analysis of the heats of formation 
(not shown) indicated that isomer {\bf 38} is not 
among the four isomers studied in Ref.\cite{ref1}, although the 
authors did not identify the isomers. Isomers with low 
relative energies include isomer {\bf 39} ($D_{5d}$, 10.9 kcal/mol) and isomer {\bf 31} 
($C_{s}$, 16.1 kcal/mol), see Fig. 1. All other isomers have 
energies higher than 20 kcal/mol. These results are in 
agreement with earlier predictions\cite{ref5,ref6}. The stabilities 
from different levels of theory agree with each other.

In Ref. \cite{ref1}, Schultz {\sl  et al}. showed an isomer with $C_{3}$ 
symmetry as the most stable structure for $C_{50}$ substituted 
with six nitrogen atoms (Fig.4b in Ref. \cite{ref1}). Here we 
present a complete search of the N$_{6}$-substituted isomers 
within the frame of C$_{50}$:{\bf 270} ($D_{3}$). Dividing the six N 
atoms into two groups of three atoms and substituting 
different groups of carbon atoms in C$_{50}$:{\bf 270} gives 64 
possible structures. We first performed calculations at 
PM3 and B3LYP/STO-3G levels, which gave similar 
relative energies. Three structures have relative energies 
lower than 16 kcal/mol, whereas all others have relative energies 
higher than 40 kcal/mol. Further calculations at the 
B3LYP/6-31G* level gave relative energies as 0.00 ($D_{3}$), 6.78 
($C_{3}$) and 9.63 ($D_{3}$) kcal/mol, confirming results from 
lower levels. The HOMO-LUMO gaps, 2.70 eV, 2.85 eV and 
2.58 eV, respectively, are large by DFT standards, 
indicating their kinetic stability. Fig.2 shows the most 
stable structure of ${\rm C}_{50}$:{\bf 270} with $N_{6}$-substitution  
having $D_{3}$ symmetry.

In summary, quantum chemical calculations at both 
DFT and PM3 levels indicate that either a complete 
survey or a more rational selection scheme is needed to 
determine the ground state structures of  fullerenes.

\vspace{0.5cm}

\begin{center}{\Large\bf Caption of Figures}\end{center}

\vspace{0.5cm}

{\bf Fig.1}: Relative energy of isomers of fullerene ${\rm C}_{40}$. 
The isomer index follows Ref.\cite{ref3}.

\vspace{0.3cm}

{\bf Fig.2}: The most stable structure of ${\rm C}_{50}$:{\bf 270} with 
$N_{6}$-substitution  has $D_{3}$ symmetry.

\end{multicols}

\end{document}